# To use or not to use cool superconductors?

Alex Gurevich

**The high critical temperature and magnetic field in cuprates and Fe-based superconductors are not enough to assure applications at higher temperatures. Making these superconductors useful involves complex and expensive technologies to address many conflicting physics and materials requirements.**

Discovered in 1911 [1], superconductivity remained a mystery for nearly 40 years until the Bardeen-Cooper-Schrieffer (BCS) theory explained its main puzzling features such as vanishing electric resistivity and the expulsion of the magnetic flux as consequences of a transition to a phase-coherent state of overlapping Cooper pairs of electrons with antiparallel spins in the s-wave orbital state glued together by lattice vibrations (phonons) [2]. The subsequent discoveries of vortices - fluxon tubes (typically 80-400 nm in diameter) of circulating currents carrying the quantized magnetic flux $\phi_0 = 2 \times 10^{-15}$ Wb packed into a hexagonal lattice - in type II superconductors, and of the Josephson effect had finalized the modern view of conventional superconductivity [3] by the mid 1960s. Triumphant confirmations of the BCS theory and discoveries of many superconducting materials quickly followed, resulting in first applications in magnetic resonance imaging (MRI), research magnets and SQUID interferometers [4,5].

By the mid 1980s, the transition temperature $T_c$ and the upper critical field $H_{c2}$, above which superconductivity disappears, were regarded as the main parameters of merit for magnet applications of superconductors, with $T_c$ between 9 and 23 K and $H_{c2}$ up to 60 T ($PbMo_6S_8$) at the liquid helium temperature, 4.2 K. The strategy of making superconductors useful seemed clear: take a material with high $T_c$ and $H_{c2}$, alloy it with nonmagnetic impurities to enhance $H_{c2}$, and then develop a technology to incorporate materials defects such as nonsuperconducting precipitates or grain boundaries to pin superconducting vortices. Pinning is necessary to prevent dissipative motion of vortices under the action of a flowing current, which otherwise would cause electric resistance even below $H_{c2}$ and $T_c$. Because stronger pinning allows a superconductor to carry larger non-dissipative current densities, up to a critical value $J_c(T,H)$ at high magnetic fields, materials optimization involved incorporating as many pinning centers as possible to maximize $J_c$ without significant degradation of $T_c$ and $H_{c2}$. Finally, composite wires were produced by embedding thin superconducting filaments into a flexible metallic (Cu) matrix to provide thermal quench stability and good mechanical properties [4]. This approach works for most conventional superconductors, in particular, NbTi ($T_c = 9$ K) and Nb$_3$Sn ($T_c = 18$ K) used in MRI magnets.

The discoveries of heavy fermions, organic superconductors, and the Chevrel phases [6] gave first indications that the conventional approach may break down for superconductors with small (nanoscale) coherence length $\xi = \hbar v_F/2\pi k_B T_c$, non s-wave symmetry of the Cooper pairs, and strong vortex fluctuations in quasi-one dimensional or layered materials. Here $\xi$ quantifies the size of the Cooper pair, $v_F$ is the Fermi velocity, $2\pi\hbar$ is the Planck constant, and $k_B$ is the Boltzman constant. These features of unconventional superconductors, first regarded as exotic and not relevant to practical conductors, were eventually recognized as being among the key issues for applications at 77 K, triggered by the groundbreaking discovery of superconductivity in the cuprates by Bednortz and Muller in 1986 [5]. At that time, the initial enthusiasm about powerful high-field magnets, motors, generators and transmission lines working at liquid nitrogen temperatures (77 K) was based on a belief that the high values of $T_c$ for YBa$_2$Cu$_3$O$_7$ ($T_c = 90$ K) and (Bi,Pb)$_2$Sr$_2$Ca$_2$Cu$_3$O$_{8+x}$ ($T_c$ up to $108$ K) would assure high-field conductors similar to those for conventional superconductors.

The reality turned out to be more complicated, but eventually (Bi,Pb)$_2$Sr$_2$Ca$_2$Cu$_3$O$_{8+x}$, Bi$_2$Sr$_2$CaCu$_2$O$_{8+x}$ and YBa$_2$Cu$_3$O$_7$ conductors have been developed [4,5]. These conductors have been used to build power distribution cables and fault current limiters which are already incorporated in the US electric grid, ship propulsion superconducting motors, and high field research magnets [7]. However, more than two decades of unprecedented R&D of the cuprates have shown that applications at 77 K are much more challenging than at 4.2 K, regardless of the values of $T_c$ and $H_{c2}$ [4,5]. Unfortunately, it appears that the materials properties responsible for high $T_c$ and $H_{c2}$ of the cuprates and the recently discovered Fe-based superconductors (FBS) [8] are also responsible for the obstacles for applications. Making such materials useful inevitably involves compromises between many conflicting requirements, defining the parameters of merit depending on the operating conditions and on the specific application.

## SIMILARITIES OF CUPRATES AND FBS

Both cuprates and FBS are layered materials in which superconductivity occurs primarily on atomic planes (the Cu-O planes in the cuprates and the Fe-As or Fe planes in FBS). FBS comprise 4 main families: ReFeAsO$_{1-x}$F$_x$ with Re = La, Sm, Nd and $T_c$ up to 55 K, (Ba$_x$M$_{1-x}$)Fe$_2$As$_2$ with M = Co, K and $T_c$ up to 38 K, MFeAs with M = Li, Na and $T_c$ up to 18 K, and FeSe$_{1-x}$Te$_x$ with $T_c$ up to 16 K [9,10]. Superconductivity in the cuprates occurs upon doping a Mott antiferromagnetic (AF) insulator while FBS become superconducting by doping a parent AF semi-metal, which means that in both cases superconductivity competes with AF states. In addition to unconventional and not yet fully understood microscopic mechanisms of superconductivity, the cuprates and FBS have many remarkable similarities which also cause problems for applications:

1. High normal state resistivities, low carrier densities, and low Fermi energies as compared to conventional superconductors. As a result, both FBS and the cuprates have small Cooper pairs with $\xi \approx$ 1-2 nm, but large Thomas-Fermi screening lengths $l_{TF} \sim \xi$.

2. Proximity of superconductivity to competing AF states.

3. Unconventional symmetry of the Cooper pairs: d-wave in cuprates and multiband s-wave with a possible sign change of the superconducting gap on disconnected pieces of the Fermi surface in FBS (the so-called s$^{\pm}$ pairing) [11].

4. Large ratios $\gamma = m_c/m_{ab}$ of the electron masses along the c-axis and the ab-plane, with $\gamma$ ranging from ~ 1 to ~ 50 for FBS and from ~ 20 to > $10^4$-$10^5$ for the cuprates.

5. Complex chemical compositions, precipitation of second phases and sensitivity of superconducting properties to local nonstoichiometry.

The points 1 and 4 contribute to the enhancement of thermal fluctuations which significantly reduce the useful T-H domain in which superconductors can carry currents even in the presence of strong pinning. The points 1, 2, 3 and 5 are responsible for current-limiting grain boundaries which deteriorate current-carrying capability of superconducting wires. To ameliorate the consequences of these materials features, complex and expensive technologies had to be developed to enable applications involving the cuprates at 77 K, and will likely have an impact on those involving FBSs.

**PINNING OF FLUCTUATING VORTICES**

Whether superconductors are used in motors, power lines, and even more so in magnets, the supercurrent flows in the presence of a magnetic field. The materials must have high critical current densities $J_c(T,H)$ preferably also for magnetic fields higher than 5T. For the cuprates, $J_c$ has been recently pushed almost to the fundamental limit by incorporating arrays of oxide (for example, $Y_2O_3$ or $BaZrO_3$) nanoparticles into $YBa_2Cu_3O_{7-x}$ films. Such pinning nanostructures can be tuned by varying the shape, size and spatial correlations of oblate or prolate nanoprecipitates, self-assembled chains of nanoparticles or nanorods, typically spaced by 4-10 nm and being 2-4 nm in diameter [12-16]. These artificial pinning centers do enhance $J_c$, particularly at low field where $J_c(77\ K, 0\ T) \sim (3\text{-}5) \times 10^{10}$ A m$^{-2}$ in thin $YBa_2Cu_3O_{7-x}$ films can approach 10%-20% of the depairing current density, the maximum supercurrent density in the absence of vortices above which the Cooper pairs break. Such "designer" pinning nanostructures also increase $J_c$ at intermediate fields most relevant to magnets. For instance, the high values $J_c(0,77K) = 2.7 \times 10^{10}$ A m$^{-2}$ and $J_c(5\ T, 77\ K) = 10^9$ A m$^{-2}$ were observed on $YBa_2Cu_3O_{7-x}$ films with $Y_2O_3$ nanoprecipitates [13]. Meanwhile, the first reports about high values of $J_c > 4 \times 10^{10}$ A m$^{-2}$ at 4.2 K [17] and $J_c$ enhanced by incorporating oxide nanopillars in $BaFe_2As_2$ films [18] indicate that strong pinning in FBS can also be achieved.

Strong pinning is only one necessary condition for applications; the field dependence of $J_c(T,H)$ is also of major importance. In conventional superconductors $J_c(T,H)$ decreases as $T$ and $H$ increase, vanishing at the field close to $H_{c2}(T)$. The value $J_c(4.2\ K, 5\ T) \sim 5 \times 10^9$ A m$^{-2}$ is characteristic of Nb47wt%Ti alloys with $T_c = 9$ K and $H_{c2}(4.2\ K) = 12$ T used in magnets [4]. Many superconductors have $H_{c2}$ much higher than in NbTi because they have shorter coherence lengths and can sustain stronger fields up to $H_{c2}(0) \sim \phi_0/2\pi\xi^2$ at which the spacing between vortices $(H_{c2}/\phi_0)^{1/2}$ becomes of the order of the diameter of nonsuperconducting vortex cores $\sim 2\xi$. Thus, very high $H_{c2}$ of the cuprates and FBS result from their short coherence lengths $\xi = \hbar v_F/2\pi k_B T_c$, either due to high $T_c \sim 90\text{-}130$ K in the cuprates or small $v_F$ in semi-metallic FBS. The values of $H_{c2}$ at $T = 0$ estimated by extrapolating low-field measurements are often well above 100 T and exceed the BCS paramagnetic limit $H_p$ at which the Zeeman energy equals the binding energy of the Cooper pair, $H_p[T] = 1.84T_c[K]$. The remarkable resilience of FBS to high magnetic fields is enhanced by the interplay of orbital and paramagnetic pairbreaking and by the multiband $s^{\pm}$ pairing [19].

Unfortunately, strong pinning and high $H_{c2}$ do not automatically make cuprates and FBS good for high-field magnets because their materials features enhance thermal fluctuations of vortices and result in current-blocking grain boundaries - a bad combination for applications. Thermal fluctuations weaken pinning and cause melting of solid vortex structures, giving rise to dissipation well below $H_{c2}$. As a result, a superconductor can carry currents without dissipation only in a smaller part of its $T$-$H$ phase diagram limited by the so-called irreversibility field $H^*(T)$ at which $J_c(H)$ vanishes. At higher fields $H^* < H < H_{c2}$ both cuprates and FBS behave as poor metals and the high values of $H_{c2}$ become irrelevant. The dissipative field domain $H^* < H < H_{c2}$ widens significantly in anisotropic materials with low carrier density, as illustrated in Figure 1 which shows that, for highly anisotropic cuprates, $H^*(T)$ can be much lower than $H_{c2}(T)$.

For the cuprates and to a lesser extent for FSB, $H^*$ is controlled by thermal fluctuations which cause melting of the vortex lattice at $H_m(T) \sim 0.005H_{c2}(0)(T_c/T - 1)^2/Gi \approx H^*$ [20]. Here the Ginzburg number $Gi = 2\gamma(k_BT_cm_{ab}/\pi\hbar^2n\xi)^2$ – the squared ratio of the thermal energy $k_BT_c$ to the superconducting condensation energy in the volume of the Cooper pair, quantifies the strength of thermal fluctuations, and $n$ is the carrier density. The vortex melting field $H^* \propto H_{c2}(0)Gi^{-1} \propto \gamma(n/m_{ab}T_c)^2$ thus decreases dramatically in anisotropic materials with high $T_c$ and low $n$. In conventional superconductors with $Gi \sim 10^{-10} – 10^{-6}$ fluctuations are negligible and $H^* \approx H_{c2}$ [20]. However, the moderately anisotropic YBa$_2$Cu$_3$O$_{7-x}$ with $Gi \sim 10^{-2}$ has $H^*(77 K) \approx 0.5H_{c2}(77 K)$, while the extremely anisotropic (Bi,Pb)$_2$Sr$_2$Ca$_2$Cu$_3$O$_x$ with $Gi \sim 1$ has $H^*(77 K) << H_{c2}$ (77 K). Different FSB have $Gi$ ranging from $\sim 10^{-4}$-$10^{-5}$ to $\sim 10^{-2}$ [10], yet even the anisotropic NdFeAsO$_{1-x}$F$_x$ with $Gi \sim 10^{-2}$, has $H^*(T) > 30$ T at temperatures $\sim$ 20-30 K where FBS have advantages over (Bi,Pb)$_2$Sr$_2$Ca$_2$Cu$_3$O$_x$ and MgB$_2$ (see Figure 1).

One may think that pinning nanostructures, which enhance $J_c$ so effectively, could also increase $H^*$ by preventing thermal wandering of vortices in the layered cuprates. However, this seems to be not the case: even for YBa$_2$Cu$_3$O$_7$ films with highest $J_c$ values, $H^*$ is not increased significantly [12-16], indicating that $H^*$ may be mostly limited by intrinsic materials parameters. It is the reduced irreversibility fields of the cuprates, which are behind one of the fundamental obstacles for high-field applications at 77 K.

**WEAKLY LINKED GRAIN BOUNDARIES**

Aside from strong vortex pinning and high $H^*$, a fundamental ingredient for successful applications is the fabrication of long polycrystalline wires. One of the main issues for the cuprates is that grain boundaries between misoriented crystallites impede current flow because the critical current density through a grain boundary $J_{gb}(\theta) = J_0 exp(-\theta/\theta_0)$ drops exponentially as the misorientation angle $\theta$ increases [21]. Here $\theta_0 \approx 3\text{-}5^o$ so the spread of misorientation angles $\Delta\theta \sim 40^o$ in polycrystals can reduce $J_{gb}$ by 2-3 orders of magnitude. Recent experiments revealed similar weak linked grain boundaries in $Ba(Fe_{1-x}Co_x)_2As_2$ bicrystals [22] and polycrystalline $LaFeAsO_{1-x}F_x$ [23].

Discovered in 1988, the current-limiting grain boundaries in cuprates [21] were immediately recognized as a serious obstacle for applications because, instead of flowing along the wire, current breaks into disconnected loops circulating in the grains as shown in Figure 2. This problem has been eventually addressed by the coated conductor technology in which the fraction of high-angle grain boundaries with $\theta > 5\text{-}7^o$ is reduced by growing $YBa_2Cu_3O_{7-x}$ films on textured metallic substrates [5]. Figure 3 shows an example of the coated conductor architecture, which makes the idea of $YBa_2Cu_3O_{7-x}$ "single crystal by the mile" a reality. Currently long (hundreds of meters) coated conductor tapes are commercially available for power and magnet applications [4,5]. Although an impressive tour de force of materials science and engineering, the coated conductors have several drawbacks: 1. Growing long $YBa_2Cu_3O_{7-x}$ films and complex buffer layers on textured substrates is much slower and more expensive than the production of conventional round multifilamentary wires, 2. Planar coated conductor geometry strongly increases the electromagnetic losses in alternating magnetic fields. 3. As evident from Figure 3, the current-carrying $YBa_2Cu_3O_{7-x}$ film takes only 1-2% of the conductor cross section. Thus, to make such conductors competitive, $J_c$ of $YBa_2Cu_3O_{7-x}$ film has to be pushed to its limit, for example, by the oxide nanoparticles.

In the case of FBSs, the first SmFeAsO$_{1-x}$F$_x$ [24,25], Sr$_{0.6}$K$_{0.4}$Fe$_2$As$_2$ [26] and FeSeTe [27] wires made by the *ex situ* powder-in-tube method exhibited poor grain connectivity and had rather low $J_c(5\ K,\ 1\ T) \approx 10^4 - 10^7$ A m$^{-2}$ because the grain boundaries in these wires are likely covered by nonsuperconducting second phases [10]. Assuming that further technological refinements will eliminate chemical granularity, porosity and other extrinsic factors, the question remains if even clean grain boundaries in FBS are intrinsic weak links as they are in the cuprates. So far clean current-limiting grain boundaries have only been observed on Ba(Fe$_{1-x}$Co$_x$)$_2$As$_2$ bicrystals [22] so it is still too early to conclude if the intrinsic weak links are indeed characteristic of all FBS. However, if they are, magnet applications of FBS may require coated conductors in some form with $J_c$ of the FBS films enhanced by pinning nanostructures.

It is natural to wonder whether the grain boundary problem may be caused by those same mechanisms which provide high values of $T_c$ and $H_{c2}$ in the cuprates and FBS. Low carrier densities, short coherence lengths, unconventional pairing symmetry, large screening lengths and competing AF states characteristic of both cuprates and FBS do contribute to the suppression of superconductivity on grain boundaries [21,28,29]. Competition of superconductivity and antiferromagnetism can cause precipitation of the parent AF phase around dislocation cores blocking current through the grain boundary [28], which can be reduced by overdoping the boundary with Ca [21,30]. In the cuprates, local nonstoichiometry, charge and strain-driven impurity segregation, and precipitation of non-superconducting and magnetic second phases at grain boundaries can also inhibit grain connectivity [30]. Much work needs to be done to understand the extent to which all these factors contribute to the weak link behavior of grain boundaries in FBS. By contrast, grain boundaries in conventional superconductors or even in two-band MgB$_2$ with $T_c = 40$ K but $l_{TF} << \xi$ are not weak links [4].

**IS ONE BETTER THAN THE OTHER?**

Common materials features of cuprates and FBS discussed here define a complex set of requirements for applications. For magnets, these requirements are best satisfied by the least anisotropic cuprate YBa$_2$Cu$_3$O$_{7-x}$ for which the coated conductor technology had to be developed to eliminate high-angle grain boundaries. The price is high: coated conductors utilize only a few percent of the current-carrying cross section, so $J_c$ of $YBa_2Cu_3O_{7-x}$ had to be pushed to its limit. Yet this is currently the only enabling technology for magnet applications at 77 K.

How do FBS look in comparison? Obviously, the lower values of $T_c < 55$ K limit applications of FBS to temperatures ~ 10 - 30 K where cryocoolers are effective. At these temperatures FBS such as $(Ba_xK_{1-x})Fe_2As_2$ or $NdFeAsO_{1-x}F_x$ have very high $H^*$ and $H_{c2}$ up to 50 T. Moreover, $H_{c2}$ of $(Ba_xK_{1-x})Fe_2As_2$ is much less anisotropic than $H_{c2}$ of $YBa_2Cu_3O_{7-x}$ and $NdFeAsO_{1-x}F_x$, which is also useful for magnets. The extent to which poor grain connectivity in FBS polycrystals is intrinsic remains unclear, not least because of the huge number of different FBS to be tested. Grain boundaries for some FSB may be more transparent to current than for others, so extensive measurements of $J_{gb}(\theta)$ on bicrystals of different FBS will be needed. If these experiments will show that the application-relevant FSB do exhibit the grain boundary problem, a coated conductor technology may be required. Interesting suggestions on managing grain boundaries in FBS films by template engineering have already been reported [17,31]. As far as pinning is concerned, thin film and single crystal FBS can carry high current densities ~ $10^{10}$ A m$^{-2}$ at 4.2 K [17,32], making them competitive with conventional superconductors.

Are FSB worth the effort then? As far as the fundamental research is concerned, FBS are exiting new materials with a rich interplay of superconductivity and magnetism. Investigation of the FBS physics may bring more surprises, perhaps giving clues for a better understanding of the cuprate superconductivity or even discovery of new high-$T_c$ materials. As for applications, it would be premature to jump to definite conclusions at this early stage. Materials properties of FBS such as high $T_c$, $H_{c2}$ and $J_c$ indicate good possibilities for magnet applications at 20-40 K, provided that the grain connectivity problem is resolved. For instance, the arsenic-free FeSeTe has a weakly anisotropic $H_{c2}$ much higher than $H_{c2}$ of $Nb_3Sn$ and $MgB_2$ at 4.2K (see Figure 1), but very low carrier densities and short $\xi$ of chalcogenides [9,10] can make them prone to poor grain connectivity. In turn, $Ba_xK_{1-x}Fe_2As_2$ may be a good magnet material because its high and moderately anisotropic $H_{c2}(20K) \approx 45$ T could make it more useful than the more anisotropic $NdFeAsO_{1-x}F_x$ with higher $T_c$. However, attractive materials properties are only one necessary prerequisite for applications; another one is the price and complexity of the conductor technology. It is therefore not surprising that so far cheaper, less anisotropic and more technological materials like NbTi, $Nb_3Sn$, $MgB_2$ or $YBa_2Cu_3O_{7-x}$ have won the race for magnet applications although many superconductors have much higher $T_c$ and $H_{c2}$. We just need to wait a bit longer to know if one of the FBS members can join this exclusive club.

*Alex Gurevich (gurevich@odu.edu) is at the Department of Physics, Old Dominion University, Norfolk, VA 23529, USA.*

**References:**

1.  Onnes, K., Comm. Physical Lab., Univ. of Leiden, Suppl. **34b** to **133-144**, 37 (1913).
2.  Bardeen, J., Cooper, L.N. & Schrieffer, J.R., Phys. Rev. **108**, 1175-1204 (1957).
3.  Tinkham, M. Introduction to Superconductivity, Mc-Grow Hill Inc. New York, London, Tokyo, 1996.
4.  Larbalestier, D., Gurevich, A., Feldmann, D.M. & Polyanskii, A, Nature **414**, 368-377 (2001).
5.  *Basic Research Needs for Superconductivity*, Report on the Basic Energy Sciences Workshop on Superconductivity May 2006, *http://www.sc.doe.gov/bes/reports/files/SC_rpt.pdf*
6.  Maple, M.B., Bauer, E.D., Zapf, V.S. & Wosnitza, J. Unconventional superconductivity in novel materials. In Superconductivity, vol. 1. Conventional and unconventional superconductors, edited by Bennemann, K.H. & Ketterson, J.D. Springer-Verlag, Berlin, Heidelberg, pp. 639-744, 2008.
7.  Malozemoff, A.P., Nature Materials **6**, 617-619 (2007).
8.  Kamihara, Y., Watanabe, T., Hirano, M. & Hosono, H., J. Am. Chem. Soc. **130**, 3296-3297 (2008).
9.  Johnston, D.C., Adv. Phys. **59**, 803-1061 (2010).
10. Putti, M. *et al.,* Supercond. Sci. Technol. **23**, 034003 (2010).
11. Mazin, I.I. & Schmalian, J., Physica C **469**, 614-629 (2009)
12. Haugan, T.J., Barnes, P.N., Wheeler, R., Meisenkothen, F. & Sumption, M.D., Nature **430**, 867-870 (2004).
13. Mele, P. *et al.,* Supercond. Sci. Technol. **19**, 44-50 (2006).
14. Kang, S. *et al.,* Science **311**, 1911-1914 (2006).
15. Gutierrez, J. *et al,.* Nature Materials **6**, 367-373 (2007).
16. Maiorov, B. *et al.,* Nature Materials **8**, 398-404 (2009).
17. Katase, T., Hiramatsu, H., Kamiya, T. & Hosono, H., Appl. Phys. Express **3**, 063101 (2010).
18. Zhang, Y. *et al.,* Appl. Phys. Lett. **98**, 042509 (2011).
19. Gurevich, A., Phys. Rev. B **82**, 184504 (2010).
20. Blatter, G., Feigelman, M.V., Geshkenbein, V.B., Larkin, A.I. & Vinokur, V.M., Rev. Mod. Phys. **66**, 1125-1388 (1994).
21. Hilgenkamp, H. & Mannhart, J, Rev. Mod. Phys. **74**, 485-549 (2002).
22. Lee, S. *et al.,* Appl. Phys. Lett. **95**, 212505 (2009).
23. Heindl, S. *et al*., Phys. Rev. Lett. **104**, 077001 (2010).
24. Gao, Z. *et al.,* Supercond. Sci. Technol. **21**, 112001 (2008).
25. Zhang, X. *et al.,* Physica C **470**, 104-108 (2010).
26. Qi, Y. *et al.,* Supercond. Sci. Technol. **23**, 055009 (2010).
27. Mizuguchi, Y. *et al.* Appl. Phys. Express **2**, 083004 (2009).
28. Gurevich, A. & Pashitskii, E.A., Phys. Rev. B**57**, 13878-13893 (1998).
29. Graser, S. *et al.,* Nature Physics **6**, 609-612 (2010).
30. Song, X., Daniels, G., Feldmann, D.M., Gurevich, A. & Larbalestier, D.C., Nature Materials **4**, 470-475 (2005).
31. Lee, S. *et al.,* Nature Materials **9**, 397-401 (2010).
32. Moll, P.J.W. *et al.,* Nature Materials **9**, 628-633 (2010).
33. Braithwaite, D. *et al.,* J. Phys. Soc. Japan **79**, 053703 (2010).
34. Jaroszynski, J. *et al.,* Phys. Rev. B **78**, 174523 (2008).
35. Altarawneh, M.M. *et al.,* Phys. Rev. B **78**, 220202(R) (2008).
36. Braccini, V. *et al.,* Phys. Rev. B **71**, 012504 (2005).
37. Chen, Bo *et al*., Nature Physics **3,** 239-242 (2007).
38. Ayai, N. *et al*., Physica C **468**, 1747-1752 (2008).
39. Baily, S.A. *et al.,* Phys. Rev. Lett. **100**, 027004 (2008).

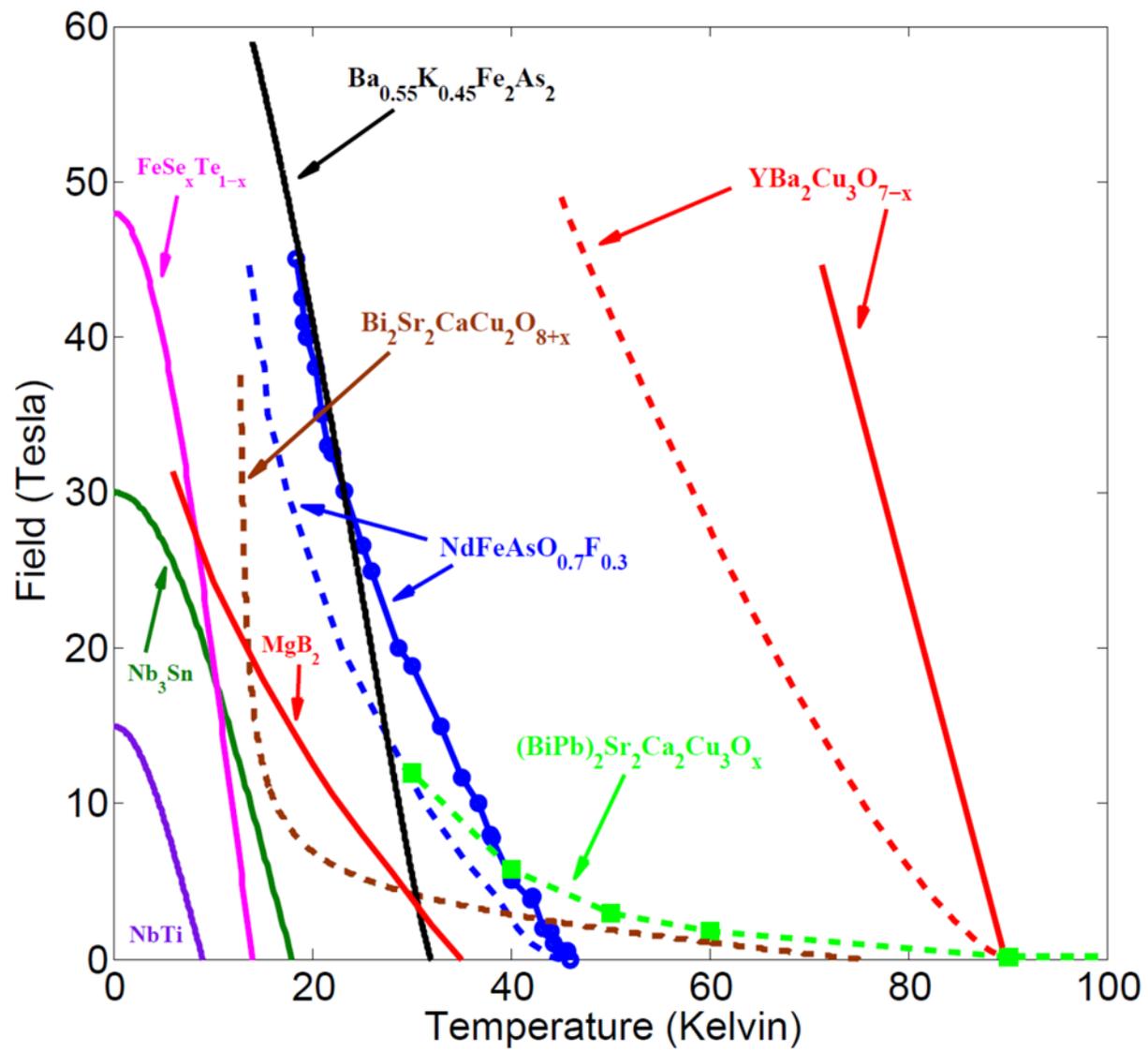

# Figure 1

Comparative T-H phase diagram for different superconducting materials. Here the solid and dashed lines show the upper critical fields $H_{c2}(T)$ and the irreversibility fields $H^*(T)$, respectively for $H//c$. Shown here are the representative members of the main FBS families: FeSeTe [33], $NdFeAsO_{0.7}F_{0.3}$ [34] and $Ba_{0.55}K_{0.45}Fe_2As_2$ [35] single crystals. The FSB are compared with the conventional NbTi, $Nb_3Sn$ [4] and $MgB_2$ [36], along with the application-relevant cuprates: the highly anisotropic $Bi_2Sr_2CaCu_2O_{8+x}$ single crystal [37], $(BiPb)_2Sr_2Ca_2Cu_3O_x$ tapes [38] used in low-temperature high field applications or low field applications at 77K (current leads, transmission lines etc), and $YBa_2Cu_3O_{7-x}$ single crystals [39]. In the most interesting for magnets range of fields, 5 T < H < 50 T, most FBS have $H_{c2}(T)$ clustered between $H^*(T)$ for the layered $Bi_2Sr_2CaCu_2O_{8+x}$ and $H^*(T)$ for the least anisotropic $YBa_2Cu_3O_{7-x}$. Both $NdFeAsO_{0.7}F_{0.3}$ and $Ba_{0.55}K_{0.45}Fe_2As_2$ could in principle provide fields up to 40-50 T at 20K. The difference between $H^*$ and $H_{c2}$ for $NdFeAsO_{0.7}F_{0.3}$ at 20-30 K is not as big as the difference between $H^*$ and $H_{c2}$ for $YBa_2Cu_3O_{7-x}$ at 77K, which reflects the diminishing role of thermal fluctuations of vortices at lower $T$. Because the less anisotropic $Ba_{0.55}K_{0.45}Fe_2As_2$ with $1 < \gamma(T) < 2$ and $T_c = 32$ K has a much higher slope $dH_{c2}/dT$ and a smaller difference between $H_{c2}$ and $H^*$ than $NdFeAsO_{0.7}F_{0.3}$ with $\gamma(T) \approx 4-8$ and $T_c = 42$ K, the conductors based on $Ba_{0.55}K_{0.45}Fe_2As_2$ could also generate fields up to 40-50 T at 20K. These data only illustrate the application potential of FBS given that further improvements of the high-field properties of FBS are likely.

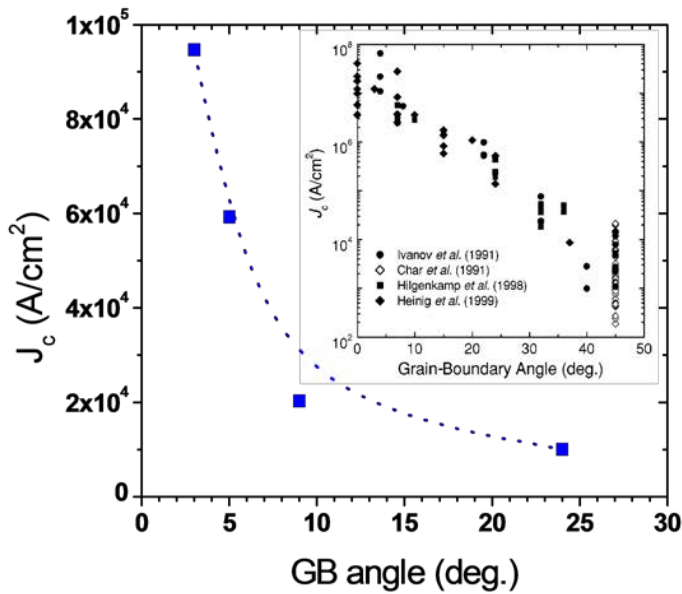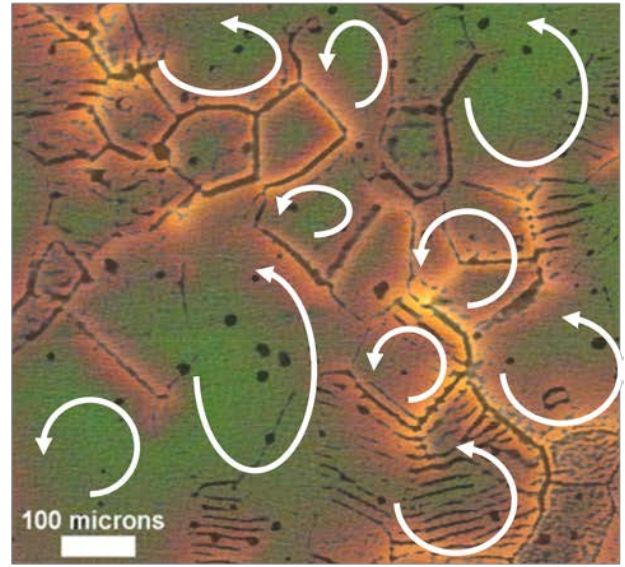

Figure 2

Weak link behavior of grain boundaries in FBS and the cuprates. Left: The dependence of the critical current density $J_c$ through the grain boundary on the misorientation angle $\theta$ in a 350 nm thick epitaxial $(Ba_{0.84}Co_{0.16})Fe_2As_2$ thin film bicrystal grown on [001] tilt (001) $SrTiO_3$ substrate [22]. Inset shows the exponential decrease of $J_c(\theta)$ in $YBa_2Cu_3O_{7-x}$ bicrystals [21]. The angular dependence of $J_c$ in $(Ba_{1-x}Co_x)Fe_2As_2$ appears similar to $J_c(\theta)$ in $YBa_2Cu_3O_{7-x}$, yielding a drop in $J_c$ by ~ 10 times as the angle increases from zero to ~ $25^0$ (this is only a qualitative conclusion given the significant scattering of $J_c(\theta)$ for different $YBa_2Cu_3O_{7-x}$ bicrystals). Right: Magnetic granularity in a $YBa_2Cu_3O_{7-x}$ polycrystal revealed by magneto-optical imaging [4]. Here the yellow contrast shows the preferential penetration of magnetic flux along the network of grain boundaries (black lines). As a result, instead of flowing along the conductor, current breaks into weakly connected current loops (white arrows) circulating in the grains. Left panel, reproduced with permission from ref. 22. Inset reproduced with permission from ref. 21. Right panel, modified from reference [4]

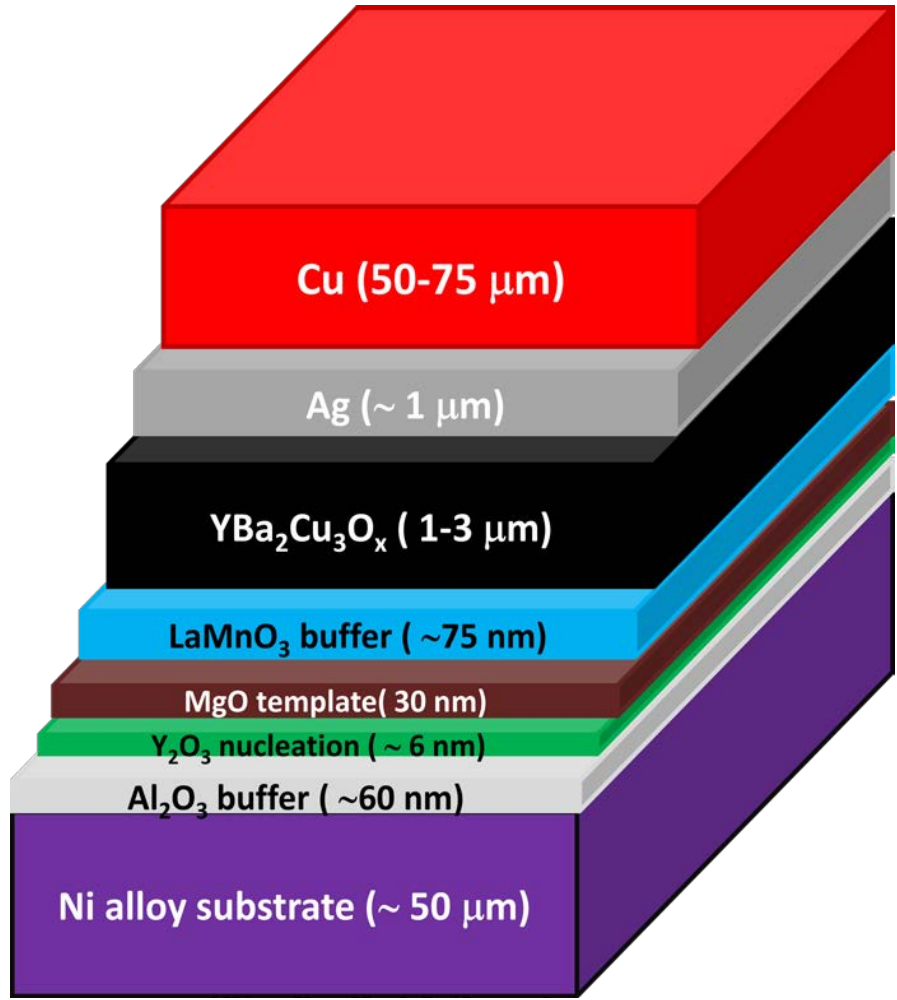

Figure 3

A typical architecture of a coated conductor made by the ion beam assisted deposition (IBAD) [5]. The $YBa_2Cu_3O_{7-x}$ film is grown on a textured Ni alloyed substrate with a complex buffer layer structure, which enables replication of the low-angle grain structure of the substrate in $YBa_2Cu_3O_{7-x}$ and protects it from chemical contamination. The stabilizing layers of Ag and Cu on top of the $YBa_2Cu_3O_{7-x}$ film provide thermal quench protection of the tape conductors which are usually few mm wide and 0.1-0.2 mm thick. Notice that the current-carrying superconducting $YBa_2Cu_3O_{7-x}$ film takes only 1-2% of the conductor cross-section, which strongly reduces the engineering current density $J_e = J_c d/D$ where $d$ = 1-3 μm is the thickness of $YBa_2Cu_3O_{7-x}$ and D = 0.1-0.2 mm is the thickness of the conductor.